\documentclass[12pt]{article}
\input tcilatex
\usepackage{amsfonts}
\author{R.M.Mir-Kasimov
\thanks{This work is partly supported by INTAS grant 93-127}
\thanks{e-mail:$<mirkr@thsun1.jinr.dubna.su>$ } \\
JINR Dubna Russia\\
and \\
Yildiz Technical University Istanbul Turkey \\
                                       }
\title{Fock's theory of hydrogen atom and quantum space}

\QQQ{Keywords}{
geometry, hydrogen atom, quantum geometry, momentum space,
curvature, representation }

\date{}
\begin{document}

\maketitle
\begin{abstract}
It is shown that V.A.Fock's theory of hydrogen atom gives an example
of the non-relativistic Snyder-like Quantum Geometry (QG).
\end{abstract}

\section{Introducton}

V.A.Fock's theory of hydrogen atom \cite{fock}
is a very bright landmark
in the history of quantum theory.
Many physicists trust that the meaning of this theory is much more
general and important than the explanation of the symmetry of one specific
atomic system though very important one.
They believe that its actual meaning
still must be understood,
that it contains some "signal form the future",
some features of the true quantum theory of particle interactions.
The search for this more general sense
of \cite{fock} can only be based on attempts to
develop V.A.Fock's idea in this or in that way, making
"experiments" on it,
for example, generalizing it to the case of the
relativistic quantum (field) theory.
And attempts to interpret \cite{fock} in a new sense
take place indeed.
Works based on the historical paper \cite{fock}
have permanently appeared since the time it was published in 1935,
certifying
the faith  of the authors
in its more profound destination.

In the present
paper an attempt has been made to look at the
V.A.Fock's approach from the point of view of Quantum
Geometry and noncommutative differential calculus.
This work is a development of the article \cite{71}.

There are no grounds to transfer the geometric notions derived from the
macroscopic experience to small (microscopic) distances.
>From the  philosophical point of view this question has
been discussed since ancient times.
The term
"$\alpha \mu \epsilon \rho \omega \sigma $"
has been introduced similtaneously with
"$\alpha \tau \omega \mu \omega \sigma $"
and denotes the smallest, indivisible portion of the space.

The suggestions to consider geometry as a subject of quantisation
natually appeared almost similtaneously with quantum theory
itself.
So Quantum Geometry (QG) is as old as quantum
theory itself.
One of the first QG models was suggested by H.Snyder
\cite{1}, \cite{2}.
In this approach
usual commuting position operators were changed for
the noncommuting quantities of a concrete form.
W.Pauli \cite{2} stressed that new Snyder's coordinate operators
could be considered as boosts of
the momentum space
of constant curvature (De Sitter or Anti De Siter momentum space).
In other words, the Snyder
quantization of space-time  is based on the
substitution of the pseudo-Euclidean geometry of the momentum
space by the De Sitter geometry.

To be more exact, actually Snyder didn't use
the connection of his quantum coordinates with
generators of the isometry group of the curved momentum space
(see \cite{2}).
The idea to consider the momentum space with
non-Euclidean  geometry as a cornerstone of the theory
with quantum space-time belongs to Yu.A.Golfand  \cite{5}
and was developed by I.E.Tamm \cite{4} and others \cite{6}-\cite{11}.

Such a change of the geometry of the momentum space leads to the
mo\-di\-fication of the procedure of extension of the S-matrix off the mass
shell \cite{9}, i.e. to the different dynamical description.
In fact, the statement on the geometry of the momentum
space off the mass shell
is an additional axiom of quantum field theory (QT).
Actually in the standard QFT, this axiom is
accepted without saying.
In the nonrelativistic theory, the extension off the
energy shell on he ground of Schr\"{o}dinger equation
in the momentum space and Lippmann-Schwinger equation
must be considered.

We can think that some background interaction exists,
which modifies the geometry of the momentum space.
See in this connection \cite{7}.
As we have stressed above, all
other axioms are fulfilled, including the standard translation invariance.
The last means that there are relative coordinates
(properly defined) which
are the subjects of quantization \cite{9}.
In  consequence of the
change of the geometry of the p-space the space-time becomes quantum
(noncommutative).

We stress that physical meaning of the geometry and
topology of the momentum space has not obtained clear physical
interpretation yet.
The space-time groups considered in QFT as the covariance groups
are the groups isometry groups of space-time.

It is worth to mentioning a series of recent papers
\cite{deser}-\cite{t'H1} where it has been
shown that the curved momentum space and the corresponding
Snyder-like quantum space naturally arise when considering
the 2+1 model of gravity interacting with the scalar field.
(The canonical momenta belong to the hyperboloid in four-dimensional
projective hyperboloid.)

The explicit character of Snyder's approach to space-time
quantization has a remarkable consequence:
we can define the spectrum of a commutative set of
operators constructed from
$\stackrel{\wedge }{x}_\mu $
and other generators of the De Sitter group.
As it has been shown in \cite{9}, the
formulation of the generalized causality condition
and QFT in terms of the
points of this new numerical quantum space-time is as comprehensive
procedure as it is in the usual QFT with the Minkovskian space-time.
In this
approach the structure of the singular field-theoretic
functions is entirely
reconstructed as compared to the standard QFT,
and the corresponding
perturbation theory is free of ultraviolet divergences.

In the present paper, we shall use the V.A.Fock's
theory of hydrogen atom symmetry to show that it is
in fact the realization of the picture described
above in the non-relativistic case:
The Coulomb field fulfills the role of
the background interaction mentioned above, which provides the noneuclidean
geometry of momentum space.
The modified shifts of the last (which are up to
some similarity transformation the Runge-Lenz vector's components) can be
considered as nonrelativistic analogs of Snyder's coordinates (\ref
{koor2}).

The paper is organized as fllows.
In Section 2, we recall briefly the
necessary moments of the V.A.Fock's theory of hydrogen atom.
In Section
3, we consider Snyder's theory.
Section 4 is devoted to the analysis of
the spectrum and matrix elements of unitary irreducible representations
of the isometry group of momentum space and the interpretation of the
spectrum as  quantum space (QS) is given.
It is shown that the
Schr\"{o}dinger equation in QS is a differential-difference
equation with the increment equal to Bohr's radius.
In Section 5, an introduction to
the noncommutative differential calculus with impact to apply it to
the Schr\"{o}dinger equation in QS is given.
Section 6 contains the theory of Schr\"{o}dinger equation in QS as a
noncommutative differential equation in QS.
In the last 7th Section an
example of integrable case (q-oscillator) of the generalised
Schr\"{o}dinger equation is given.

\section{Snyder approach}

In this approach
the usual quantum mechanical coordinate operators

\begin{equation}
\label{koor1}
x_{\mu}=i\hbar \frac \partial {\partial p^{\mu} },\qquad \mu
=0,1,2,3\qquad g_{\mu \nu }=\mbox{diag}(1,-1,-1,-1)
\end{equation}

the generators of translations of the Minkovskian momentum space are
substituted by Snyder quantum coordinates
$\stackrel{\wedge }{x}_\mu $
i.e. the generators of De Sitter boosts:

\begin{equation}
\label{koor2}
\stackrel{\wedge }{x}_{\mu} =
il_{0}\left( p_{4}
\frac {\partial}{\partial p^{\mu} }-
p_{\mu} \frac {\partial} {\partial p^{4}}\right),
\qquad
\left[
\stackrel{\wedge }{x}_{\mu},\stackrel{\wedge }{x}_{\nu} \right] =
-il_{0}^{2}
\stackrel{\wedge }{M}_{\mu \nu }
\end{equation}

where $l_{0}$ , the ''fundamental length'', indicates the scale
at which the effects of QG become appreciable.
A plausible candidate for this
role is the Planck's length
$l_{Planck}=\sqrt{\frac{c\hbar }{G}}$.

\section{Fock's theory of Hydrogen atom}

The Schr\"odinger equation for Hydrogen atom (we limit ourselves with the
case of continuum part of spectrum)

\begin{equation}
\label{schr1}
H\psi (\stackrel{\rightarrow }{x})=\left( E_p-\frac{e^2}{\left|
\stackrel{\rightarrow }{x}\right| }\right) \psi (\stackrel{\rightarrow }{x}%
)=E_q\psi (\stackrel{\rightarrow }{x})=\frac{q^2}{2\mu }\psi (\stackrel{%
\rightarrow }{x})
\end{equation}

where
$E_p=\frac{p^2}{2\mu },\ E_q=\frac{q^2}{2\mu }$,
in the momentum
representation takes the form

\begin{equation}
\label{schr2}
\left( p^2-q^2\right) \psi (\stackrel{\rightarrow }{p})=
\frac
\hbar {\pi ^2a}\int \frac{d^3p^{\prime }\
\psi (\stackrel{\rightarrow }
{p^{\prime }})}{\left|
\stackrel{\rightarrow }{p^{\prime }}-\stackrel{
\rightarrow }{p}\right| ^{2}}
\end{equation}

where $a$ is the atomic unit of length (Bohr's radius)

\begin{equation}
\label{at1}
a=\frac{\hbar ^{2}}{\mu e^{2}}
\end{equation}

Recall that atomic units of energy,
momentum and time are correspondingly

\begin{equation}
\label{at2}
e_{a}=
\frac{\mu e^{4}}{\hbar ^{2}}
,\qquad
\pi _{a}=
\frac{\mu e^{2}}\hbar
,\qquad
\tau _{a}=
\frac{\hbar ^{3}}{\mu e^{4}}
\end{equation}

Following V.A.Fock \cite{fock} we take into account the
non-euclidean geometry of momentum space, naturally arising here.
For the
continuous part of the energy spectrum it is pseudoeuclidean 3 dimensional
space of negative curvature (Lobachevsky space).
Introducing 4-dimensional projective momenta
$P_{\mu} \left( \mu =0,1,2,3\right) $ we have

\begin{equation}
\label{ksi1}
\stackrel{\rightarrow }{P}=
\frac{2q\stackrel{\rightarrow }{p}}
{\stackrel{\rightarrow }{p}^{2}-q^{2}}
\quad
P_{0}=
\frac{\stackrel{\rightarrow }{p}^{2}+q^2}
{\stackrel{\rightarrow }{p}^2-q^{2}}
\quad
q=\sqrt{2\mu E}\quad
\end{equation}

\begin{equation}
\label{hyp1}
P_0^2-\stackrel{\rightarrow }{P}^2=
1\
\end{equation}

Or inverse relation

\begin{equation}
\label{inv1}
\stackrel{\rightarrow }{p}=
\frac{q\stackrel{\rightarrow }{P}}
{P_{0}-1}
\quad
p^{2}=q^{2}\frac{P_{0}+1} {P_{0}-1}
\end{equation}

The equation (\ref{hyp1}) describes the two-sheet hyperboloid
(the upper
sheet corresponds to
$1\leq P_0<+\infty $,
the lower pole to
$-\infty <P_{0}\leq -1$ ).

It is convenient to use the hyperspherical coordinates

\vspace*{0.5cm}
$$
\stackrel{\rightarrow }{P}=
\sinh \alpha \stackrel{\rightarrow }{n}\quad
P_{0}=\pm \cosh \alpha
$$
\begin{equation}
\label{ksi2}
\end{equation}
$$
\stackrel{\rightarrow }{n}=
\frac{\stackrel{
\rightarrow }{P}}{\left| \stackrel{\rightarrow }{P}\right| }=
\left( \sin \theta \cos \phi ,\sin \theta \sin \phi ,\cos \theta \right)
$$
\vspace*{0.5cm}

The volume element in the Lobachevsky space is

\begin{equation}
\label{vol1}
d\Omega _P=
\frac{d^3P}{\left| P_0\right| }=
\sinh {}^2\alpha \sin
\theta d\alpha d\theta d\phi
\end{equation}

The distance
$s(P,P^{\prime })$
of two points of the Lobachevsky space (\ref
{hyp1}) is given by

\begin{equation}
\label{dist1}
\cosh s(P,P^{\prime })=
\left| 1-\frac{\left( P-P^{\prime
}\right) ^2}2\right|
\end{equation}

In terms of
$P_{\mu}$
the equation (\ref{schr2}) takes the form

\begin{equation}
\label{schr3}
\Phi _{r}(P)=
\frac r{2\pi ^2a}\int \frac{d\Omega _{P^{\prime }}}
{\left( P-P^{\prime }\right) ^{2}}
\Phi _{r}(P^{\prime })
\end{equation}

where

\begin{equation}
\label{r1}
\Phi _{r}(P)=(p^{2}-q^{2})^{2}\psi (\stackrel{\rightarrow }{p}),\,
r=\frac{\hbar}{q}
\end{equation}

This equation is manifestly invariant under the group of motions of the
Lobachevsky momentum space which is Lorentz group. The generators of Lorentz
group boosts

\begin{equation}
\label{runge1}
\stackrel{\wedge }{x}_i=
-i\left[ P_0\frac \partial {\partial
P^i}-P_i\frac \partial {\partial P^0}\right]
\end{equation}

up to some similarity transformation coincide with the
additional integrals
of motion of the Coulomb problem i.e. Runge-Lenz invariants.
>From the other
side their similarity to Snyder coordinates (\ref{koor2})
is evident.

\section{Wave functions and differential-dif\-fe\-ren\-ce
Schr\"{o}\-din\-ger equa\-ti\-on
in quan\-tum space}

As it was shown by V.Fock \cite{fock} the solutions of the Schr\"odinger
equation in momentum space (\ref{schr3}) are the eigen-functions of the
Laplace-Beltrami operator on the Lobachevsky space (\ref{hyp1}),
or the Casimir operator of the Lorentz group:

\begin{equation}
\label{cas1}\left( \stackrel{\rightarrow }{x}^2-\frac 1{\hbar ^2}\stackrel{%
\rightarrow }{L}^2\right) \Phi _r(P)=\left( 1+\frac{r^2}{a^2}\right) \Phi
_r(P)
\end{equation}

where
$\stackrel{\rightarrow }{L}$
is the vector of angular momentum operators.
The solutions of (\ref{cas1}) are the matrix elements of unitary
irreducible (infinite dimensional) representations of the Lorentz group.
For the principal series of unitary representations of
Lorentz group the
parameter
$r$
runs over the interval

\begin{equation}
\label{interv1}
0\leq r<\infty
\end{equation}

which coincides
with the physically admissible region of variation (see (\ref{r1})).
>From the other side the interval (\ref{interv1}) is the range of variation
of the relative distance in three-dimensional Euclidean space.
Taking these properties of $r$ into account we
interpret the parameter
$r$
as the relative distance in the quantum space
\cite{7}, \cite{8}.

Let us consider the quantities

\begin{equation}
\label{plw1}
\Phi _r(P)=
<\stackrel{\rightarrow }{r}\mid \stackrel{\rightarrow
}{P}>=\left| P_0-\stackrel{\rightarrow }{P}\stackrel{\rightarrow }{n_{r}}
\right| ^{-1-i\frac {r}{a}}
\end{equation}

where

\begin{equation}
\label{plw2}
\stackrel{\rightarrow }{r}=
r\stackrel{\rightarrow }{n_r},\qquad
\stackrel{\rightarrow }{n_r}^2=1
\end{equation}

The expression (\ref{plw1}) from one hand side is the solution of the
equation (\ref{cas1}), from the other hand side it
is the generating function
for the radial solutions of the Schr\"{o}dinger
equation in the momentum space (\ref{schr3}).
The expression (\ref{plw1}) plays the role of the plane wave in
quantum r-space.
The radial solutions of the Schr\"{o}dinger equation in the
momentum space can be obtained from the expansion in spherical harmonics:

\begin{equation}
\label{sh1}
<\stackrel{\rightarrow }{r}\mid \stackrel{\rightarrow }{P}>=
4\pi\sum_{l=0}^\infty \sum_{m=-l}^{m=l}i^l<\rho,l,\mid \alpha ><\stackrel
{\rightarrow }{n_r}\mid l,m><l,m\mid \stackrel{\rightarrow }{n}>
\end{equation}

where
$\rho$
is the dimensionless parameter:

\begin{equation}
\label{rho1}
\rho=\frac{r}{a}
\end{equation}

and

\begin{equation}
\label{sh2}
<l,m\mid \stackrel{\rightarrow }{n}>=
<l,m\mid \theta ,\phi
>=Y_{lm}(\stackrel{\rightarrow }{n}),
\quad
<\stackrel{\rightarrow }{n}\mid l,m>=
Y_{lm}^{*}(\stackrel{\rightarrow }{n})
\end{equation}

The functions
$<\rho,l\mid \alpha >$
are the radial wave functions of the
Coulomb problem in the momentum space.
They can be obtained also as the Fourier transforms of the
radial solutions in configurational space.
We present the different
representations of
$<\rho,l\mid \alpha >$
in terms of Legendre functions
$P_{i\rho -\frac 12}^{-l-\frac 12}\left( \cosh \alpha \right) $ :

\begin{equation}
\label{sh3}
<\rho,l\mid \alpha >=(-i)^l\sqrt{\frac \pi {2\sinh \alpha }}\frac
{\Gamma (i\rho +l+1)}
{\Gamma (i\rho +1)}P_{i\rho -\frac 12}^{-l-\frac
12}\left( \cosh \alpha \right)
\end{equation}

Gegenbauer functions
$C_{i\rho -l-1}^{l+1}\left( \cosh \alpha \right) $ :

\vspace*{0.5cm}
$$
<\rho,l\mid \alpha >=
$$
\begin{equation}
\label{sh4}
\end{equation}
$$
(-i)^l\sqrt{\frac \pi {2\sinh \alpha }}\left(
\frac{\sinh \alpha }2\right) ^{l+\frac 12}\frac{\Gamma \left( 2l+2\right)
\Gamma (i\rho -l)}{\Gamma (i\rho +1)\Gamma (l+\frac 32)}C_{i\rho
-l-1}^{l+1}\left( \cosh \alpha \right)
$$
\vspace*{0.5cm}

and hypergeometric function
$_2F_1\left( \alpha ,\beta ;\gamma ;z\right) $ :

\vspace*{0.5cm}
$$
<\rho,l\mid \alpha >=(-i)^l
\sqrt{\frac \pi {2\sinh \alpha }}\left( \frac{\sinh \alpha }2\right)
^{l+\frac 12}\frac{\Gamma (i\rho +l+1)}
{\Gamma (i\rho +1)\Gamma (l+\frac {3}{2})}
$$
\begin{equation}
\label{sh5}
\end{equation}
$$
_2F_1\left( i\rho +l+1,-i\rho +l+1;l+\frac 32;-\sinh {}^2\frac \alpha
2\right)=
$$
\vspace*{0.5cm}

$$
=(-i)^l
\sqrt{\frac \pi {2\sinh \alpha }}\left( \frac{\sinh \alpha }2\right)
^{l+\frac 12}\frac{\Gamma (i\rho +l+1)}{\Gamma (i\rho +1)
\Gamma (l+\frac 32)}
e^{\alpha \left( i\rho -l-1\right) }
$$
\begin{equation}
\label{sh6}
\end{equation}
$$
\ _2F_1\left( -i\rho +l+1,l+1;2l+2;2e^{-\alpha }\sinh \alpha \right)=
$$
\vspace*{1.5cm}

$$
=
\sqrt{\frac \pi 2}\left( -i\frac{\sinh \alpha }2\right) ^l\frac{\Gamma
(i\rho +l+1)}{\Gamma (i\rho +1)\Gamma (l+\frac 32)}
$$
\begin{equation}
\label{sh7}
\end{equation}
$$
_2F_1\left( \frac{i\rho +l+1}2,\frac{-i\rho +l+1}2;l+\frac 32;-\sinh
{}^2\alpha \right)
$$
\vspace*{0.5cm}

These different representations are convenient for performing the
contraction limit when we consider in the next section the correspondence
with usual (nonquantum) space limit.
The following orthogonality and
completeness conditions for the radial solutions are valid
\vspace*{0.5cm}

\begin{equation}
\label{sh8}
\frac 2\pi \int_0^\infty \sinh {}^2\alpha \ d\alpha <\rho ,l\mid
\alpha ><\alpha \mid \rho ^{\prime },l>=
\frac{\delta \left( \rho -\rho
^{\prime }\right) }{\rho ^2}
\end{equation}
\vspace*{0.5cm}
\begin{equation}
\label{sh9}\frac 2\pi \int_0^\infty \rho ^2d\rho <\alpha \mid \rho ,l><\rho
,l\mid \alpha ^{\prime }>=\frac{\delta \left( \alpha -\alpha ^{\prime
}\right) }{\sinh {}^2\alpha }
\end{equation}
\vspace*{0.5cm}

and corresponding conditions for the plane waves (\ref{plw1}):
\vspace*{0.5cm}

\begin{equation}
\label{sh10}
\frac 1{\left( 2\pi \right) ^3}\int <\stackrel{\rightarrow }{r}
\mid \stackrel{\rightarrow }{P}><\stackrel{\rightarrow }{P}
\mid \stackrel
{\rightarrow }{r^{\prime }}>d\Omega _P=
\delta \left( \stackrel{\rightarrow }
{r}-\stackrel{\rightarrow }{r^{\prime }}\right)
\end{equation}
\vspace*{0.5cm}

\begin{equation}
\label{sh11}
\frac 1{\left( 2\pi \right) ^3}
\int <\stackrel{\rightarrow }{P}
\mid
\stackrel{\rightarrow }{r}><\stackrel{\rightarrow }{r}\mid \stackrel
{\rightarrow }{P^{\prime }}>d^3r=
\delta \left( \stackrel{\rightarrow }{P}-
\stackrel{\rightarrow }{P^{\prime }}\right) P_0
\end{equation}
\vspace*{0.5cm}

The plane wave in quantum space and its radial part obey the following
equations off the energy shell i.e. for
$E_p\neq E_q$ or $p\neq q$ .
\vspace*{0.5cm}

$$
e_{a} \left[ \cosh \left( i\frac \partial {\partial \rho}\right) +
\frac{i}\rho\sinh \left( i\frac \partial {\partial \rho}\right) -
\frac{\Delta
_{\theta ,\phi }}{\rho^2}e^{i\frac \partial {\partial \rho}}-
1\right] <\stackrel
{\rightarrow }{r}\mid \stackrel{\rightarrow }{P}>=
$$
\begin{equation}
\label{dschr1}
\end{equation}
$$
\stackrel{\wedge }{H}_0
<\stackrel{\rightarrow }{r}
\mid
\stackrel{\rightarrow}{P}>=
E_P<\stackrel{\rightarrow }{r}\mid \stackrel{\rightarrow }{P}>
$$

\vspace*{0.5cm}

$$
e_{a}\left[ \cosh \left( i\frac \partial {\partial \rho}\right) +
\frac{i}{\rho}\sinh \left( i\frac \partial {\partial \rho}\right) -
\frac{l(l+1)}{\rho^2}
e^{i\frac \partial {\partial \rho}}-
1\right] <\rho ,l\mid \alpha >=
$$
\begin{equation}
\label{dschr2}
\end{equation}
$$
=\left( \stackrel{\wedge }{H}_{0l}-E_P\right) <\rho ,l\mid \alpha >
$$

where
$E_P=e_{a}\left( \left| P_0\right| -1\right) =
2e_{a}\sinh {}^2\frac \alpha
2
,\quad
e_{a}=\frac{\mu e^4}{\hbar ^2}$
is the atomic unit of energy.
The strong argument for the idea that plane wave
$<\stackrel{\rightarrow }{r}\mid
\stackrel{\rightarrow }{P}>$
describes the free motion in the quantum
$r$
-space is the existence of more three differential-difference operators
$
\stackrel{\wedge }{p_i}$
for which
$<\stackrel{\rightarrow }{r}\mid \stackrel
{\rightarrow }{P}>$
is the eigenfunction with eigenvalues equal to the
momentum components

\begin{equation}
\label{comp1}
\stackrel{\wedge }{p}_i<\stackrel{\rightarrow }{r}\mid
\stackrel{\rightarrow }{P}>=p_i<\stackrel{\rightarrow }{r}\mid \stackrel
{\rightarrow }{P}>
\end{equation}

where

\vspace*{0.5cm}
$$
\stackrel{\wedge }{p}_1=
\pi _{a} \left\{ \sin \theta \cos \phi
\left( e^{i\frac \partial {\partial \rho}}-\stackrel{\wedge }{H}_0\right)
-i\left( \frac{\cos \theta \cos \phi }\rho
\frac \partial {\partial \theta }-
\frac{\sin \phi }{\rho\sin \theta }\frac \partial {\partial \phi }\right)
e^{i\frac \partial {\partial \rho}}\right\}
$$
\begin{equation}
\label{comp2}
\stackrel{\wedge }{p}_2=
\pi _{a}
\left\{ \sin \theta \sin \phi
\left( e^{i\rho\frac \partial {\partial \rho}}-\stackrel{\wedge }{H}_0\right)
-i\left( \frac{\cos \theta \sin \phi }\rho\frac \partial {\partial \theta }+
\frac{\cos \phi }{\rho\sin \theta }\frac \partial {\partial \phi }\right)
e^{i\frac \partial {\partial \rho}}\right\}
\end{equation}
$$
\stackrel{\wedge }{p}_3=
\pi _{a}
\left\{ -\cos \theta \left(
e^{i\frac \partial {\partial \rho}}-\stackrel{\wedge }{H}_0\right) +
i\frac{\sin \theta }{\rho}\frac
\partial {\partial \theta }e^{i\frac \partial {\partial
\rho}}\right\}
$$
\vspace*{0.5cm}

$
\pi _{a}=\frac{\mu e^{2}}{\hbar}
$
is the atomic unit of momentum.

It looks now  quite  natural to make the next
step and introduce the interaction term
$V(r)$
into the free differential-difference Schr\"{o}dinger equation in
quantum space (\ref{dschr1}).
>From the usual point of view this corresponds
to some perturbations for the Coulomb potential.
We stress that there are
even integrable cases for such differential-difference
equations (see \cite{12} and
references therein).

\section{Contraction}

The important requirement to the theory with curved momentum space is its
correspondence with the usual theory.
In the physical regime, when we can
neglect the effects of curvature all relations must go over into the usual
ones.
Let us first analyse this problem in the momentum space (\ref{ksi1}).
The vicinities of the tops of both poles

\begin{equation}
\label{top1}
P_0\approx \pm 1
\end{equation}

carry the flat geometry.
For these regions we are in the regime when the
In\"{o}n\"{u}-Wigner contraction (\cite{Inonu}) is actual approach.
For example
the Snyder's quantum coordinate operators (\ref{runge1}) go over into usual
coordinate operators (\ref{koor1}) in these regions.

The tops of the hyperboloid (\ref{hyp1}) are
\vspace*{0.5cm}

\begin{enumerate}

\vspace*{0.5cm}
\item

\begin{equation}
\label{top11}
P_0=1\ or\ q\cong 0,\, i.e.
\, \rho  \gg
\frac{\pi _a }{\left|\stackrel{\rightarrow }{p}\right| }
\end{equation}
\vspace*{0.5cm}

\item

\vspace*{0.5cm}

\begin{equation}
\label{top2}
P_0=-1\ or\ q\rightarrow \infty ,\ i.e.\, \, \left| \stackrel
{\rightarrow }{p}\right|\ll \frac{\pi _a }{\rho }
\end{equation}
\vspace*{0.5cm}

\end{enumerate}

In classical physics the small
$p$-s correspond to great impact parameters.
In
this regime the scattered particle slightly feels the Coulomb field.
In the
case of the bound states we must consider the orbits corresponding to big
values of principal quantum number
$n$.
In the contraction limit all
finite-difference relations reduce to standard differential relations of
Quantum Mechanics.
For example the differential-difference operators of
momentum (\ref{comp2}) reduce to usual momentum operators

\begin{equation}
\label{comp3}
\stackrel{\wedge }{p}_ie^{\frac{i\stackrel{\rightarrow }{r}
\stackrel{\rightarrow }{P}}\hbar }=
-i\hbar \frac \partial {\partial x_i}e^
{\frac{i\stackrel{\rightarrow }{r}\stackrel{\rightarrow }{p}}\hbar }=
p_ie^{
\frac{i\stackrel{\rightarrow }{r}\stackrel{\rightarrow }{p}}\hbar }
\end{equation}

and the plane wave (\ref{plw1}) converts to usual exponential function:

\vspace*{0.5cm}
$$
\left| P_0-\stackrel{\rightarrow }{P}\stackrel{\rightarrow }{n_{r}}\right|
^{-1-i\frac ra}
$$

$$
<\stackrel{\rightarrow }{r}\mid \stackrel{\rightarrow }{P}>=
\exp \left\{
-\left( 1+i\frac ra\right) \ln \left( P_0-\stackrel{\rightarrow }{P}
\stackrel{\rightarrow }{n_r}\right) \right\} \approx
$$
\begin{equation}
\label{con1}
\end{equation}
$$
\approx \exp \left\{ -\left( 1+i\frac ra\right) \ln \left( 1-
\stackrel{\rightarrow }{P}\stackrel{\rightarrow }{n_r}+\cdot \cdot \cdot
\right) \right\} \approx
$$

$$
\approx \exp \left\{ i\frac ra\stackrel{\rightarrow }{P}\stackrel
{\rightarrow }{n_r}+\cdot \cdot \cdot \right\}
\approx e^{i\frac{\stackrel
{\rightarrow }{P}\stackrel{\rightarrow }{x}}\hbar }
$$
\vspace*{0.5cm}

\section
{Non-Commutative differential calculus and finite-difference derivatives}

We start this section with an historical remark.
Referring to the
second Snyder's paper on quantum space (the second paper of \cite{1})
we invite the reader to convince that his generalisation of
Maxwell's equations
for the case of QG is based, in fact, on a version of the
noncommutative differential calculus.

Let us show that the finite-difference Schr\"odinger equation (\ref{dschr2})
is na\-tu\-rally described in terms of noncommutative differential calculus
\cite{Woron}-\cite{Muller}.
This calculus can be naturally and most easily
introduced on a ground of the theory of differential forms as its
deformation. We shall limit ourselves with the differential calculus over
the associative algebra $A$ over $\mathbb{R}$ or $\mathbb{C}$.
In our case
the necessity to consider an algebra over $\mathbb{C}$ follows from the form
of finite-difference Schr\"odinger equation, containing shifts by the
imaginary quantity $ia$.
This is general property of finite-difference
Schr\"odinger equation (\ref{dschr2}) corresponding to the continuous part
of the spectrum of hydrogen atom, requiring to consider the wave functions
in the complex
$\rho $-plane.
Finite linear combinations of elements of $A$
and finite products are again elements of $A$.
The multiplication is
associative.
A differential calculus on $A$ is a $\mathbb{Z}$- graded
associative algebra over $\mathbb{C}$

\begin{equation}
\label{df1}
\Omega \left( A\right) =
\sum_{r=0}\ _{\oplus }\ \Omega ^r\left(
A\right)
\end{equation}

\begin{equation}
\label{df11}
\Omega ^0\left( A\right) =A,\quad \Omega ^r\left( A\right)
=\{0\}\,\forall r<0
\end{equation}

The elements of $\Omega ^r\left( A\right) $ are called $r$-forms.
There
exist an exterior derivative operator $d$ which satisfies the following
conditions

\begin{equation}
\label{df2}
d^2=0
\end{equation}

and

\begin{equation}
\label{df3}
d\left( \omega \omega ^{\prime }\right) =
\left( d\omega \right)
\omega ^{\prime }+(-1)^r\omega d\omega ^{\prime }
\end{equation}

where $\omega $
and $\omega ^{\prime }$
are
$r-$
and
$r^{\prime }-$
forms,
respectively.
$A$ is the commutative algebra generated by the coordinate
functions $x^i,\ i=1,...n$. In the standard differential calculus on usual
manifolds differentials commute with functions:

\begin{equation}
\label{df4}\left[ x^i,\ dx^j\right] =0,\quad i,j=1,...n
\end{equation}

in terms of real coordinates $x^i$ .
For us it is essential that (\ref{df4})
can be ge\-ne\-ra\-li\-sed (deformed) in different ways
with (\ref{df1}-\ref{df3})
still true.
Let us consider in more detail the deformation of (\ref{df4}) of
the form

\begin{equation}
\label{df5}
\left[ x^i,\ x^j\right] =0
\end{equation}

\begin{equation}
\label{df6}
\left[ x^i,\ dx^j\right] =\sum_{k=1}^ndx^kC^{ij}{}_k
\end{equation}

where the $C^{ij}{}_k$ are (complex)
constants which are constrained by the
requirement of a consistent differential calculus.

\begin{itemize}
\item
Let us apply $d$ to (\ref{df5}) and use (\ref{df6}), this gives

$$
d\left[ x^i,\ x^j\right] =-\left[ x^j,\ d\ x^i\right] +\left[
x^i,\ d\ x^j\right] =
$$
\begin{equation}
\label{df61}
\end{equation}
$$
=-\sum_{k=1}^ndx^kC^{ji}{}_k+\sum_{k=1}^ndx^kC^{ij}{}_k=0
$$

or

\begin{equation}
\label{df7}
C^{ij}{}_k=C^{ji}{}_k
\end{equation}

This means in particular

\begin{equation}
\label{df71}
\left[ x^i,\ dx^j\right] =
\left[ x^j,\ dx^i\right]
\end{equation}

The last relation can be proved directly:

\vspace{0.5cm}
$$
\left[ x^i,\ dx^j\right] =\left( dx^i\right) x^j-x^jdx^i= \\
\\
$$
\begin{equation}
\label{df72}
\end{equation}
$$
=\underbrace{d\left( x^ix^j-x^jx^i\right) }_{\mbox{=0}}+x^jdx^i-x^idx^j=
\left[ x^j,\ dx^i\right]
$$
\vspace{0.5cm}

\item
Taking the commutator of
$dx^i$
with (\ref{df5}) we obtain

$$
\left[ \left[ x^i,\ dx^j\right] ,\ dx^k\right] =
$$
$$
=x^i\left\{ dx^kx^j+\sum_{l=1}^ndx^lC^{jk}{}_l\right\} -x^j\left\{
dx^kx^i+\sum_{l=1}^ndx^lC^{ik}{}_l\right\} -
$$
$$
-dx^k\left( x^ix^j-x^jx^i\right)
=
$$
\begin{equation}
\label{df73}
\left\{ dx^kx^i+\sum_{m=1}^ndx^mC^{ik}{}_m\right\}
x^j+\sum_{l=1}^n\left\{ dx^lx^i+\sum_{m=1}^ndx^mC^{il}{}_m\right\}
C^{jk}{}_l-
\end{equation}
$$
\left\{ dx^kx^j+\sum_{m=1}^ndx^mC^{jk}{}_m\right\} x^i-\sum_{l=1}^n\left\{
dx^lx^j+\sum_{m=1}^ndx^mC^{jl}{}_m\right\} C^{ik}{}_l-
$$
$$
dx^k\left( x^ix^j-x^jx^i\right) =\sum_{l,m=1}^ndx^m\left(
C^{il}{}_mC^{jk}{}_l-C^{jl}{}_mC^{ik}{}_m\right) =0
$$

or

\begin{equation}
\label{df8}
\sum_{l=1}^n\ C^{ik}{}_lC^{jl}{}_m=\sum_{l=1}^nC^{jk}{}_l\
C^{il}{}_m
\end{equation}

using (\ref{df7}) the last equation can also be written in a form

\begin{equation}
\label{df9}
\sum_{l=1}^n\ C^{k[i}{}_lC^{j]l}{}_m=0
\end{equation}

This means that $n$ matrices $C^i$ with entries $C^{ik}{}_l$ mutually
commute.

\item
Taking the commutator of $x^k$ with (\ref{df6}) also yields
(\ref{df8}) and therefore no additional conditions.
\end{itemize}

Acting with $d$ on (\ref{df6}) and using the Leibniz rule (\ref{df3}) we
obtain the classical commutation rule

\begin{equation}
\label{df10}
dx^idx^j=-dx^jdx^i
\end{equation}

for differentials.
The equations obtained by commuting $x^k$ through these
relations are identically satisfied.

The Hodge $*$ operator (or duality transformation) for the noncommutative
differential forms is introduced by the standard formula

\begin{equation}
\label{df28}
*\left( dx^{i_1}\cdot \cdot \cdot dx^{i_k}\right) =\frac
1{\left( n-k\right) !}\sum \epsilon _{^{i_1}\cdot \cdot \cdot
^{i_ki_{k+1}}\cdot \cdot \cdot ^{i_n}}dx^{i_{k+1}}\cdot \cdot \cdot dx^{i_n}
\end{equation}

For convenience we shall make difference between right
$\stackrel{
\rightarrow }{*}$
and left
$\stackrel{\leftarrow }{*}$
Hodge operators.
By
definition
$\stackrel{\rightarrow }{*}$
acts on the forms of the type

\begin{equation}
\label{df26}
\sum \left( dx^{i_1}\cdot \cdot \cdot dx^{i_k}\ f(x)\right)
\end{equation}

$\stackrel{\leftarrow }{*}$
acts on the forms of the type

\begin{equation}
\label{df27}
\sum \left( f(x)\ dx^{i_1}\cdot \cdot \cdot dx^{i_k}\right)
\end{equation}

in both cases by the standard formula (\ref{df28}).
Action of the operator
$\stackrel{\rightarrow }{*}$
on forms of the type (\ref{df27}) and action of
the operator
$\stackrel{\leftarrow }{*}$
on the forms of the type (\ref{df26}) gives $0$ .
Correspondingly we introduce right and left
$\delta $
operations

\begin{equation}
\label{df29}
\stackrel{\rightarrow }{\delta }=\stackrel{\rightarrow }{*}\ d\
\stackrel{\rightarrow }{*},\qquad \stackrel{\leftarrow }{\delta }=\stackrel
{\leftarrow }{*}\ d\ \stackrel{\leftarrow }{*}
\end{equation}

Let $A$ be the algebra of all functions on $\mathbb{C}$.
In what follows we
consider the one dimensonal case.
It is generated by canonical coordinate
function of one variable
$\psi $$(\rho )=\rho $ .
One of the simplest
deformation of the ordinary differential calculus on $A$ is

\begin{equation}
\label{df111}
\left[ d\rho ,\ \rho \right] =\frac {i}{2}\ d \rho
\end{equation}

where $i$ (in dimensional units $ia$)
is the step in the fi\-ni\-te-dif\-fe\-ren\-ce Schr\"o\-din\-ger
eq\-ua\-tion.
To es\-tab\-lish the con\-nec\-ti\-on bet\-ween the
non\-com\-mu\-ta\-ti\-ve
dif\-fe\-ren\-ti\-al cal\-cu\-lus and
fi\-ni\-te-dif\-fe\-ren\-ce
operations in (Schr\"odinger eq.) is our goal here.
This
is a special case of of the commutation structure (\ref{df6}) considered
above.
Similar relations are encountered when considering the differential
calculus on the lattice \cite{Woron} - \cite{Muller}.
Equation (\ref{df111})
can be rewritten in a form

\begin{equation}
\label{df12}
d\rho \ \rho =\left( \rho +\frac {i}{2}\right) d\rho \
\end{equation}

which can be generalised to he total algebra $A$ as

\begin{equation}
\label{df13}
d\rho \ \psi (\rho )=\psi \left( \rho +\frac {i}{2}\right) d\rho \
\end{equation}

Then we can introduce the generalised derivatives (left and right)
corresponding to our deformed differential calculus.
For the left derivative
we write

\begin{equation}
\label{df14}
d\psi (\rho )=\left( \stackrel{\rightarrow }{\partial }\psi
(\rho )\right) \ d\rho
\end{equation}

>From Leibniz rule (\ref{df3}) we have

\begin{equation}
\label{df15}
\begin{array}{c}
d\left( \psi (\rho )\ \varphi (\rho )\right) =d\rho \ \left(
\stackrel{\rightarrow }{\partial }\left( \psi (\rho )\ \varphi (\rho
)\right) \right) =\left( d\psi (\rho )\right) \ \varphi (\rho )+\psi (\rho
)\ \left( d\varphi (\rho )\right) = \\  \\
=d\rho \ \left( \stackrel{\rightarrow }{\partial }\psi (\rho )\right) \
\varphi (\rho )+\psi (\rho )\ d\rho \ \left( \stackrel{\rightarrow }{%
\partial }\varphi (\rho )\right)
\end{array}
\end{equation}

after using (\ref{df13})

\begin{equation}
\label{df16}
d\left( \psi (\rho )\ \varphi (\rho )\right) =d\rho \ \left(
\stackrel{\rightarrow }{\partial }\psi (\rho )\right) \varphi (\rho )+d\rho
\ \psi (\rho +\frac {i}{2})
\left( \stackrel{\rightarrow }{\partial }\varphi
(\rho )\right)
\end{equation}

Now from the commutativity rule (\ref{df5})

\begin{equation}
\label{df17}
\psi (x)\ \varphi (x)=\varphi (x)\ \psi (x)
\end{equation}

it follows also that equivalent Leibniz rule is valid

\begin{equation}
\label{df18}
\begin{array}{c}
d\left( f(x)\ g(x)\right) =dx\ \left(
\stackrel{\rightarrow }{\partial }\left( f(x)\ g(x)\right) \right) =\left(
dg(x)\right) \ f(x)+g(x)\ \left( df(x)\right) = \\  \\
=dx\ \left( \stackrel{\rightarrow }{\partial }g(x)\right) \ f(x)+dx\
g(x+\frac {i}{2})\ \left( \stackrel{\rightarrow }{\partial }f(x)\right)
\end{array}
\end{equation}

Equalizing (\ref{df16}) and (\ref{df18}) we obtain

\begin{equation}
\label{df19}
\left( \stackrel{\rightarrow }{\partial }\psi (\rho )\right) \
\left[ \varphi (\rho +i\frac a2)-\varphi (\rho )\right] =
\left( \stackrel
{\rightarrow }{\partial }\varphi (\rho )\right)
\left[ \psi (\rho +\frac {I}{2})-\psi (\rho )\right]
\end{equation}

or

\begin{equation}
\label{df20}
\frac{\stackrel{\rightarrow }{\partial }\psi (\rho )}{\psi (\rho
+\frac {i}{2})-\psi (\rho )}=
\frac{\stackrel{\rightarrow }{\partial }\varphi
(\rho )}{\varphi (\rho +\frac {i}{2})-\varphi (\rho )}=const
\end{equation}

where
$const$
is the same for any function under differentiation.
To calculate this $const$ we choose

\begin{equation}
\label{df21}
\psi (\rho )=\rho
\end{equation}

This gives

\begin{equation}
\label{df22}
d\rho =d\rho \left( \stackrel{\rightarrow }{\partial }\rho
\right)
\Longrightarrow \left( \stackrel{\rightarrow }{\partial }\rho
\right) =1
\end{equation}

and

\begin{equation}
\label{df23}
const=\frac{\stackrel{\rightarrow }{\partial }\rho }{(\rho
+\frac {i}{2})-\rho }=\frac {2}{i}
\end{equation}

The ultimate expression for the left partial derivative is

\begin{equation}
\label{df24}
\stackrel{\rightarrow }{\partial }\psi (\rho )=
\frac{\psi (\rho
+\frac {i}{2})-\psi (\rho )}{\frac {i}{2}}
\end{equation}

The expression for the right derivative
$\stackrel{\leftarrow }{\partial }
\psi (\rho )$
is obtained in similar way and has a form

\begin{equation}
\label{df25}
\stackrel{\leftarrow }{\partial }\psi (\rho )=
\frac{\psi (\rho
)-\psi (\rho -\frac {i}{2})}{\frac {i}{2}}
\end{equation}

\section
{Non-commutative differential calculus and Schr\"{o}dinger equation
in quantum space}

In this section we shall apply the noncommutative differential calculus of
prevous section to Schr\"{o}dinger equation (\ref{dschr2}). Let us exclude the
''first finite-difference radial derivative'' from this equation making a
substituton
\begin{equation}
\label{sc1}\psi _l(\rho )=\frac{<\rho ,l\mid \alpha >}\rho
\end{equation}
Recall that similar substitution excludes the first radial derivative in the
usual Schr\"{o}dinger (differential) equation. Taking into account the foolowing
rules for finite-difference operations
$\sinh \left( i\frac \partial{\partial \rho }\right) $
and
$\cosh \left( i\frac \partial {\partial \rho}\right) $ :

$$
\sinh  i\frac \partial {\partial \rho } \psi (\rho )\varphi
(\rho )=
$$
\begin{equation}
\label{sc11}
\end{equation}
$$
=\sinh  i \frac {\partial}{\partial \rho } \psi (\rho )\cosh
i\frac \partial {\partial \rho } \varphi (\rho )+
\cosh
i\frac \partial {\partial \rho } \psi (\rho )\sinh  i \frac
\partial {\partial \rho } \varphi (\rho )
$$

\vspace*{0.5cm}

$$
\cosh i\frac \partial {\partial \rho } \psi (\rho )\varphi
(\rho )=
$$
\begin{equation}
\label{sc12}
\end{equation}
$$
=\cosh i\frac \partial {\partial \rho } \psi (\rho )\cosh
i\frac \partial {\partial \rho } \varphi (\rho )
+\sinh
i\frac \partial {\partial \rho }  \psi (\rho )\sinh i\frac
\partial {\partial \rho } \varphi (\rho )
$$

and relations

\begin{equation}
\label{sc13}
\cosh i\frac \partial {\partial \rho } \frac
1\rho =\frac \rho {\rho ^2+1},
\qquad
\sinh  i\frac \partial
{\partial \rho } \frac 1\rho =\frac{-i}{\rho ^2+1}
\end{equation}

We obtain

$$
e_{a}\left[ \cosh \left( i\frac \partial {\partial \rho }\right) \
-\frac{l(l+1)}{2\rho }e^{i
\frac \partial {\partial \rho }}\frac 1\rho -1+V(\rho
)-E_P\right] \psi _l(\rho )=
$$
\begin{equation}
\label{sc2}
=e_{a}\left[ 2\sinh {}^2\left(
\frac {i}{2} \frac \partial {\partial \rho }\right)
\ -\frac{l(l+1)}{2\rho }e^{i\frac \partial {\partial \rho }}\frac
1\rho +V(\rho )-E_P\right] \psi _l(\rho )=
\end{equation}
$$
=\left( \rho \stackrel{\wedge }{H}_{0l}\frac 1\rho +V(\rho )-E_P\right) \
\psi _l(\rho )=0
$$

Let us consider the expression

$$
\frac{1}{2}
\left( \stackrel{\rightarrow }{\delta }d+
\stackrel{\leftarrow }{\delta }d\right) \psi (\rho )=
\frac{1}{2}\left( \stackrel{\rightarrow }{\delta }+
\stackrel{\leftarrow }{\delta }\right) d\psi (\rho )=
\frac{1}{2}\left( \stackrel{\rightarrow }{*}d
\stackrel{\rightarrow }{*}d+
\stackrel{\leftarrow }{*}d\stackrel{\leftarrow }
{*}d\right) \psi (\rho )=\
$$
$$
=\frac 12\left( \stackrel{\rightarrow }{\delta }+
\stackrel{\leftarrow }
{\delta }\right)
\left( \left( \stackrel{\rightarrow }{\partial }\psi
(\rho)\right) d\rho +d\rho
\left( \stackrel{\leftarrow }{\partial }\psi (\rho
)\right) \right) =
$$
$$
=\frac {1}{2}\left( \stackrel{\rightarrow }{\delta }+
\stackrel{\leftarrow }{\delta }\right)
\left[ \left( \frac{\psi (\rho +\frac {i}{2})-\psi (\rho )}
{\frac {i}{2}}\right) d\rho +
d\rho \left( \frac{\psi (\rho )-\psi (\rho -
\frac{i}{2})}{\frac {i}{2}}\right) \right] =
$$
\begin{equation}
\label{df262}=
\frac 12\left( \stackrel{\rightarrow }{*}+
\stackrel{\leftarrow}
{*}\right) d\left[ \frac{\psi (\rho +\frac {i}{2})-
\psi (\rho -\frac {i}{2})}
{\frac {i}{2}}\right] =
\end{equation}
$$
-\left( \stackrel{\rightarrow }{*}+
\stackrel{\leftarrow }{*}\right)
\left[ \left( \psi (\rho +i)-\psi (\rho
)-\psi (\rho +\frac {i}{2})+
\psi (\rho -\frac {i}{2})\right) d\rho \right. +
$$
$$
\left. +d\rho \left( \psi (\rho -i)-\psi (\rho )+\psi (\rho +
\frac{i}{2})-\psi (\rho -\frac {i}{2})\right) \right] =
$$
$$
=-2 \left( \cosh i\frac \partial {\partial \rho }-1\right) \psi
(\rho )=-4\sinh {}^2\frac {i}{2}\frac \partial {\partial \rho }\
\psi (\rho )
$$

Using this result we can define left and right momentum operators in a form

\begin{equation}
\label{sc3}
\stackrel{\rightarrow }{p}=-i\hbar \stackrel{\rightarrow }{*}d,
\qquad
\stackrel{\leftarrow }{p}=-i\hbar \stackrel{\leftarrow }{*}d
\end{equation}

and free particle momentum operator

\begin{equation}
\label{sc4}
\stackrel{\wedge }{p}=\frac 12\left( \stackrel{\rightarrow }{p}+
\stackrel{\leftarrow }{p}\right)
\end{equation}

where
$\stackrel{\rightarrow }{p}$
is not a vector but the right momentum
operator!
So the first term in (\ref{sc2}) takes a form

\begin{equation}
\label{sc45}
\frac{\stackrel{\wedge }{p}^2}{2\mu }=
\frac {1}{2\mu }
\left( -2\pi_0\sinh {}\frac {i}{2}
\frac \partial {\partial \rho }\right) ^2
\end{equation}

The centrifugal term in (\ref{sc2}) can be written in a form

\begin{equation}
\label{sc5}
\frac{l(l+1)}{\rho (\rho +i)}e^{i\frac \partial {\partial \rho
}}=
\frac 1\rho \stackrel{\rightarrow }{\lambda }^2\frac 1\rho
\end{equation}

where

\begin{equation}
\label{sc6}
\stackrel{\rightarrow }{\lambda }=
\sqrt{l(l+1)}\left( 1-\frac{
\stackrel{\rightarrow }{p}}{2\pi _0}\right)
\end{equation}

Ultimately we obtain  the Schr\"{o}dinger equation in
quantum space in terms of noncommutative differential calculus as

\begin{equation}
\label{sc7}
H\psi _l(\rho )=\left( \frac{\stackrel{\wedge }{p}^2}{2\mu }
+\frac 1\rho \stackrel{\rightarrow }{\lambda }^2\frac 1\rho +
V(\rho )\right) \psi _l(\rho )=E_P\psi _l(\rho )
\end{equation}

\section{Linear oscillator=$q$-oscillator in Quantum Space}

>From usual point of wiev the interaction term $V(\rho )$ in the
differential- difference Schr\"odinger equation (\ref{dschr1}) corresponds
to the perturbed Coulomb potential. Let us consider an example of integrable
case for the Schr\"{o}dinger equation with interaction.
We write the  ladder operators
$$
a^{\pm }=\mp \frac i{\sqrt{2}\pi _{a}
\cos \frac {r}{2l_{0}}}e^{\pm \frac
12\left( \frac {r}{\lambda _{0}}\right) ^2}\left( \stackrel{\wedge }{p}
\right) e^{\mp \frac 12\left( \frac r {\lambda _0}\right) ^2}
$$
\begin{equation}
\label{lad1}{}
\end{equation}
$$
=\pm \frac{i\sqrt{2}}{\cos \frac {r}{2l_0}}e^{\pm \frac 12\left( \frac
\rho {\lambda _0}\right) ^2}\left( \sinh \frac{ia}2\frac \partial {\partial
r}\right) e^{\mp \frac 12\left( \frac {r}{\lambda _0}\right) ^2}
$$

$\stackrel{\wedge }{p}$
is the non-commutative differential operator of
radial momentum  introduced in previous section,
$\omega $
is the frequency,
$\lambda _0$
is a parameter of dimension of length:

\begin{equation}
\label{lad11}
\lambda _0=\sqrt{\frac \hbar {\mu \omega }}
\end{equation}

The ladder operators (\ref{lad1}) obey the deformed commutation relation

\begin{equation}
\label{lad2}
\left[ a^{-},a^{+}\right]
_q=qa^{-}a^{+}-q^{-1}a^{+}a^{-}=2\left( q^{-1}-q\right)
\end{equation}

which guarantees the exact solubility of this finite-difference
problem, $q$ is an dimensionless quantity,
parameter of deformation, which is expressed
in terms of physical parameters:

\begin{equation}
\label{lad21}
q=e^{-\frac{a^2}{4\lambda _0^2}}=e^{-\frac{\hbar \omega }{4e_a}
}=e^{-\frac{\omega \hbar ^3}{4\mu e^4}}
\end{equation}

We introduce the Hamiltonian

\begin{equation}
\label{lad3}
\stackrel{\wedge }{H}=\frac 12\left\{ a^{-},a^{+}\right\}
_q=\frac 12\left\{ qa^{-}a^{+}+q^{-1}a^{+}a^{-}\right\}
\end{equation}

obeying deformed commutation relations with ladder operators

\begin{equation}
\label{lad31}\left[ a^{\pm },\stackrel{\wedge }{H}\right] _{q^{\mp 1}}=\pm
\left( q^2-q^{-2}\right) a^{\pm }
\end{equation}

which in fact guarantee the integrability, and obtain the energy spectrum

\begin{equation}
\label{lad4}
E_n=2e_a\left( e^{\frac{\hbar \omega }{2e_a}(n+\frac 12)}-\cosh
\frac{\hbar \omega }{4e_a}\right)
\end{equation}

This integrable case can be easily identified with well
known $q$-oscillator.

\end{document}